\documentclass[aps,prl,twocolumn,showpacs,showkeys,groupedaddress]{revtex4-1}

\usepackage{graphicx}
\usepackage[applemac]{inputenc}

\begin{document}


\title{Ordered Array of Single Adatoms with Remarkable Thermal Stability:  Au$/$Fe$_3$O$_4$(001)}




\author{Zbyn\v{e}k Novotn\'{y}, Giacomo Argentero, Zhiming Wang, Michael Schmid, Ulrike Diebold, and Gareth S. Parkinson}
\email{parkinson@iap.tuwien.ac.at}

\affiliation{Institute of Applied Physics, Vienna University of Technology, Vienna, Austria.}



\begin{abstract}

Gold deposited on the Fe$_3$O$_4$(001) surface at room temperature was studied using Scanning Tunneling Microscopy (STM) and X-ray photoelectron spectroscopy (XPS). This surface forms a ($\sqrt2$~$\times$~$\sqrt2$)R45$^{\circ}$ reconstruction, where pairs of Fe and neighboring O ions are slightly displaced laterally producing undulating rows with `narrow' and `wide' hollow sites.  At low coverages, single Au adatoms adsorb exclusively at the narrow sites, with no significant sintering up to annealing temperatures of 400 $^{\circ}$C.  We propose the strong site preference to be related to charge and orbital ordering within the first subsurface layer of Fe$_3$O$_4$(001)-($\sqrt2$~$\times$~$\sqrt2$)R45$^{\circ}$.  Due to its high thermal stability, this could prove an ideal model system for probing the chemical reactivity of single atomic species.  

\end{abstract}

\maketitle

The study of isolated adatoms on surfaces has revealed interesting physical, magnetic and chemical phenomena including quantum confinement, Kondo scattering, and charging effects \cite{Ternes}. In addition, single metal adatoms at oxide surfaces have been proposed as the active species in several catalytic reactions \cite{Fu, Qiao, Kyriakou, Fang}, in the ultimate extension of the well known size-effect \cite{Haruta}. However, this hypothesis has remained controversial because unambiguous identification of single atoms in real catalysts is not straightforward. Verification would be of both fundamental and economic importance, since loading a catalyst with single atoms would require dramatically less precious metal \cite{Fu}. 

A successful approach to understanding fundamental mechanisms in catalysis has been to create model systems containing monodisperse nanoparticles on well characterized oxide supports, and to study them under idealized conditions \cite{Freund, Judai}. Substrates exhibiting anchoring sites such as defects \cite{Zhang, Brown}, local adsorption minima within moire patterns \cite{Zhou, Berdunov, Giordano, Nilius2}, and long range reconstructions \cite{Degen, Schmid} have proven particularly well suited to the task. However, no suitable model system exists for the study of single adatom chemistry because sintering occurs well below realistic reaction temperatures \cite{Farmer, Brown}. Nevertheless, low-temperature experiments have adatom charging \cite{Brown, Giordano, Rienks, Nilius, Nilius2, Zhang, Rim, Repp}, seen as important for their purported reactivity. In this paper we describe a model system in which Au adatoms are stabilized up to 400 $^{\circ}$C in a deep adsorption well located at 2-fold coordinated “narrow” sites of the Fe$_3$O$_4$(001)-($\sqrt2$~$\times$~$\sqrt2$)R45$^{\circ}$ surface. We propose the site preference to be linked to charge and orbital ordering in the subsurface layers of the Fe$_3$O$_4$ substrate, a material exhibiting strong electron correlation. Fe$_3$O$_4$(001)-Au thus represents an ideal model system for the study of Au adatoms and their associated chemistry, and may prove useful in resolving the role of single adatoms in catalysis. 

 \begin {figure}[b]
 \includegraphics [width=3.0 in,clip] {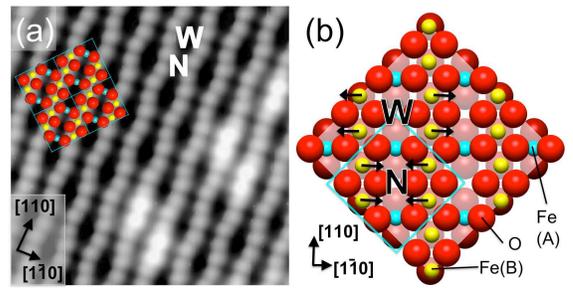}
 \caption{
(Color online) (a) STM image (60~$\times$~60~\AA{}$^2$, V$_{sample}$ = +1 V, I$_{tunnel}$ = 0.35 nA) of the clean Fe$_3$O$_4$(001) surface. The bright double protrusions located on the Fe(B) rows are due to hydroxyl species that result from dissociation of water in oxygen vacancies \cite{ParkinsonWater}.(b) Top view of the ($\sqrt2$~$\times$~$\sqrt2$)R45$^{\circ}$ reconstructed Fe$_3$O$_4$(001) surface \cite{Pentcheva2}. Alternate pairs of surface Fe(B) cations (yellow) relax perpendicular to the Fe(B) row (relaxation indicated by blue arrows), creating two distinct hollow sites within the reconstructed surface cell; wide (W) and narrow (N). The ($\sqrt2$~$\times$~$\sqrt2$)R45$^{\circ}$ unit cell is indicated, drawn with its corners in the W sites.}
\end{figure}

STM images of the Fe$_3$O$_4$(001)-($\sqrt2$~$\times$~$\sqrt2$)R45$^{\circ}$ surface exhibit undulating rows of protrusions with an average separation of 6 \AA{} (see Figure 1a) \cite{Wiesendanger, Stanka}. The undulations result from a subtle lattice distortion \cite{Pentcheva, Pentcheva2} where alternate pairs of surface Fe(B) atoms (octahedral coordination in the bulk, yellow) relax laterally by $\approx$ 0.1 \AA{} in opposite directions perpendicular to the row (for a schematic see Figure 1b). Each Fe(B) pair is located between subsurface Fe(A) cations (tetrahedral coordination, cyan). The letters N and W indicate ``narrow" and ``wide" hollow sites, respectively, where surface Fe(A) atoms would reside if the bulk structure were continued; these sites are rendered inequivalent by the reconstruction. Theory predicts that the lattice distortion is coupled to charge and orbital ordering in the subsurface layers \cite{Lodziana, Mula}, inducing a half-metal to insulator transition in the surface layers. This description is consistent with the structure of antiphase domain boundaries \cite{ParkinsonAPDB}, and scanning tunneling spectroscopy measurements, where a surface bandgap of 0.2 eV was measured \cite{Jordan}. Thus, as a substrate for nanoparticle growth, Fe$_3$O$_4$(001) can be viewed as a thin semiconducting layer on a metallic support. As such, the system bears some resemblance to the ultrathin oxide films on metals studied extensively by Freund and coworkers \cite{Giordano, Brown, Rienks, Nilius}, where charging of adatoms via tunneling has been demonstrated. 

Adsorption at the Fe$_3$O$_4$(001)-($\sqrt2$~$\times$~$\sqrt2$)R45$^{\circ}$ surface is strongly influenced by the reconstruction. Fe adatoms adsorb exclusively at the narrow sites \cite{ParkinsonFe}, while Fe dimers straddle the narrow site at higher coverage \cite{Lodziana, ParkinsonFe, Spiridis}; the wide sites remain unoccupied. Water adsorption is dissociative \cite{Kendelewicz, MulaWater, ParkinsonWater}, initially via a highly efficient reaction with oxygen vacancies \cite{MulaWater, ParkinsonWater}. This leads to pairs of H species adsorbed at narrow sites in neighboring unit cells \cite{ParkinsonWater}. The presence of the H causes the neighboring Fe(B) pair to be imaged bright \cite{ParkinsonWater, ParkinsonH}, see Figure 1a. The typical H coverage on the clean surface is 0.08 ML, i.e., 8$\%$ of the narrow sites are occupied, and no diffusion out of these sites is observed. Motivated by the strong preference for narrow site occupation observed for other adsorbates, we decided to investigate the adsorption of gold at the Fe$_3$O$_4$(001)-($\sqrt2$~$\times$~$\sqrt2$)R45$^{\circ}$ surface.

Gold at the Fe$_3$O$_4$(001) surface has been studied previously using STM and other techniques. Jordan et al. \cite{JordanAu} and Spiridis et al. \cite{SpiridisAu} reported Au cluster formation at room temperature, though there was disparity about the size of the smallest clusters. Jordan et al. interpreted their data as 2D clusters approximately 0.9 nm in diameter, with a consistent location with respect to the reconstruction. Spiridis et al. found 2D and 3D clusters. The high-resolution STM data presented here reveal that isolated Au adatoms adsorb exclusively in the narrow sites and remain there up to 400~$^{\circ}$C.

The experiments were performed using a synthetic single crystal grown via the floating zone method by Prof. Mao and coworkers at Tulane University, USA \cite{Mao}. The clean surface was prepared in an ultrahigh vacuum (UHV) vessel (base pressure under 10$^{-10}$ mbar) by cycles of Ar$^+$ sputtering (1 keV, 1.6 $\mu$A, 10 minutes), UHV annealing (700~$^{\circ}$C, 15 minutes), and annealing in O$_2$ (700~$^{\circ}$C, $p_\mathrm{O_2}$ = 2$\times$10$^{-6}$ mbar, 60 minutes). Au (99.99$\%$ purity) was deposited from a Mo crucible with a carefully degassed liquid-N$_2$ cooled e-beam evaporator (Omicron EFM3) to the surface at room temperature (RT).  In what follows, the coverage of Au is defined in terms of the narrow sites in the surface layer, i.e. 1 monolayer (ML) Au = 1 Au atom per ($\sqrt2$~$\times$~$\sqrt2$)R45$^{\circ}$ surface unit cell (1.42$\times$10$^{14}$ atoms/cm$^2$).  The deposition rate was 1 ML/min as calibrated using a water-cooled quartz crystal microbalance. During deposition a retarding voltage of +1.5 kV was applied to the cylindrical electrode in the orifice of the evaporator (flux monitor) to repel positively charged ions that are produced in the EFM3. STM and XPS measurements were performed in an adjoining analysis chamber (base pressure below 5$\times$10$^{-11}$ mbar) using a customized Omicron $\mu$-STM operated in constant current mode at room temperature with an electrochemically etched W tip. For XPS measurements, a dual anode X-ray source (VG XR3E2) and SPECS EA10+ energy analyzer were used.

 \begin {figure}[t]
 \includegraphics [width=3.0 in,clip] {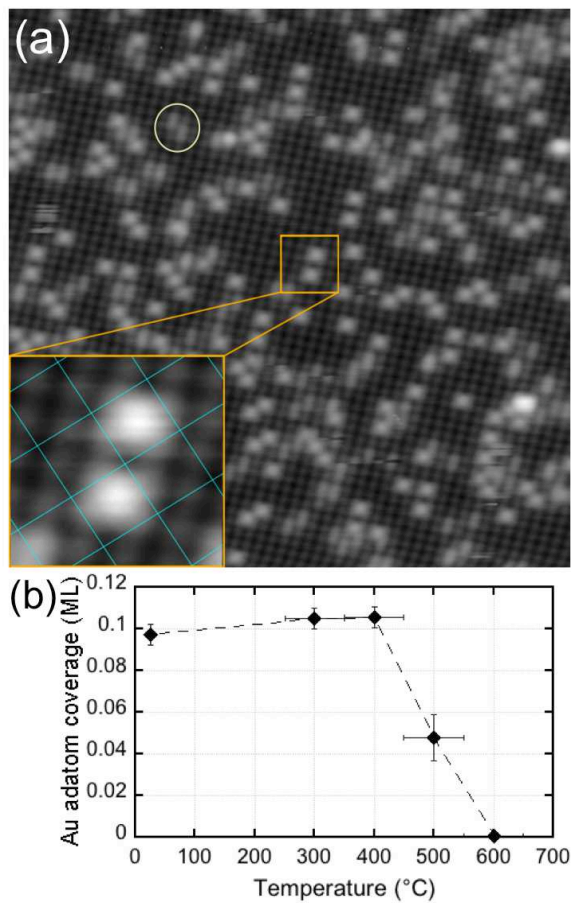}
 \caption{
(Color online) (a) STM image (300$\times$300 \AA{}$^2$, V$_{sample}$ = +1 V, I$_{tunnel}$ = 0.38 nA) of 0.12 ML Au deposited on Fe$_3$O$_4$(001) at RT. Au adatoms are located between the surface Fe(B) rows. The inset shows two Au adatoms in high resolution with the ($\sqrt2$~$\times$~$\sqrt2$)R45$^{\circ}$ periodicity indicated by cyan lines; the nodes are centered in the wide sites of the surface reconstruction. The Au adatoms are located in the center of the cell, i.e., at the narrow sites. Hydroxyl species (while oval) \cite{ParkinsonWater, ParkinsonH} have a similar coverage to the clean surface. (b) Coverage of single Au adatoms after annealing 0.1 ML Au to various temperatures.  The disappearance of Au adatoms is accompanied by the formation of Au clusters.
}
\end{figure}

In Figure 2 we present an STM image following deposition of 0.12 ML Au on the clean Fe$_3$O$_4$(001) surface at RT. Two types of protrusions are observed. Hydroxyl groups (white oval) appear with an identical coverage to that observed on the clean surface \cite{ParkinsonWater} and thus are not linked to the deposition of Au. In addition, 0.12 ML bright protrusions are located in between the Fe(B) rows. The inset shows two such protrusions. Drawing lines joining the “wide” sites of the surrounding surface we see that the protrusions are located in the center, i.e. at the narrow sites. STM movies acquired by recording several images over the same sample area show the Au adatoms to be essentially immobile at room temperature.

Figure 2b shows the density of adatoms following 0.1 ML Au deposition as a function of sample post-annealing temperature (flash anneal). Up to 400~$^{\circ}$C the coverage is constant within the statistical error. After annealing at 500~$^{\circ}$C the number of Au adatoms is 50$\%$ of the initial value, and larger Au clusters appear in the images. Annealing at 600~$^{\circ}$C transforms all adatoms into clusters.

XPS spectra for the Au 4f region for both 0.12 ML Au and a thicker Au film are shown in Figure 3. For the thicker film the Au 4f $_{7/2}$ peak position is consistent with metallic Au (83.8 eV) \cite{Torelli}, for the adatom surface it is shifted to higher binding energy by 0.6 eV, to 84.4 eV.

 \begin {figure}[t]
 \includegraphics [width=3.0 in,clip] {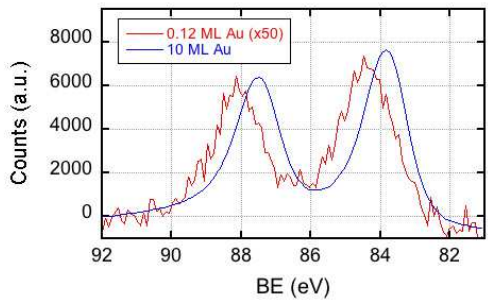}
 \caption{
(Color online) Au 4f XPS spectra for 0.12 ML Au (red) and a thicker (10 ML) Au film (blue). A spectrum of the clean surface was subtracted as a background.
}
\end{figure}

Figure 4 shows a 300$\times$300 \AA{}$^{2}$ STM image after deposition of 0.8 ML Au at RT. Isolated adatoms are again present (0.17 ML). Several locations (e.g., inside the yellow circles) show ordered Au adatom arrays with a nearest neighbor distance of 8.4 \AA{}, the periodicity of the ($\sqrt2$~$\times$~$\sqrt2$)R45$^{\circ}$ reconstruction. In addition, clusters of various sizes are observed. An analysis assuming shapes resembling truncated spheres finds that the smallest clusters contain 5 atoms. Au clusters are observed at step edges and the regular terrace (examples are labeled 1 and 2 in Figure 4, respectively). Larger clusters are often found alongside asymmetric denuded zones i.e. areas devoid of Au adatoms; the denuded zone associated with cluster 2 is encompassed by the dashed magenta outline in Figure 4. 

 \begin {figure}[t]
 \includegraphics [width=3.0 in,clip] {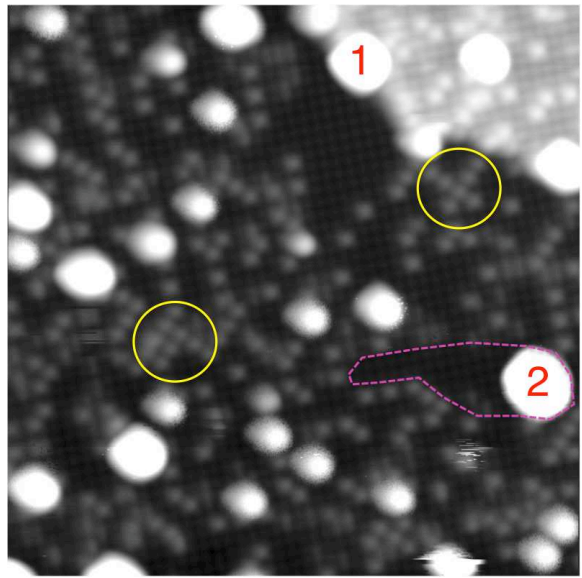}
 \caption{
(Color online) STM image (300$\times$300 \AA{}$^2$, V$_{sample}$ = +1 V, I$_{tunnel}$ = 0.31 nA) of the Fe$_3$O$_4$(001) surface with 0.8 ML Au. The yellow circles highlight areas containing a local high density of Au adatoms (One per ($\sqrt2$~$\times$~$\sqrt2$)R45$^{\circ}$ unit cell). Au nanoclusters are found at step edge sites (e.g., cluster 1) and also on the regular terrace (e.g., cluster 2). The dashed magenta line indicates an area of clean surface associated with cluster 2.
}
\end{figure}

From the distribution of Au adatoms at low coverage (Figure 2) we propose that Au atoms arrive from the gas phase, land on the surface, and diffuse a short distance to occupy the nearest narrow site, where they are strongly bound. However, as the Au coverage is increased, Au atoms will inevitably arrive at locations where all neighboring narrow sites are already filled (e.g., in the centre of the yellow circles in Figure 4). The additional Au atom interacts with one of the adatoms in a neighboring narrow site, destabilizing its bond to the surface. We do not observe dimers, trimers, or tetramers, however. Therefore we propose that a dimer is formed, and that it is less strongly bound than an adatom. Hence, as soon as it forms, it diffuses across the surface, picking up further Au adatoms. The cluster grows in size until a critical size is reached and it becomes immobile. Such a `rolling snowball' mechanism of cluster growth explains the trails of uncovered surface observed near some larger clusters (magenta dashed line in Figure 4). 

The experimental data show that the Fe$_3$O$_4$(001)-($\sqrt2$~$\times$~$\sqrt2$)R45$^{\circ}$ surface acts as an adsorption template for Au adatoms. While we cannot infer the details of the adsorption geometry from the STM images, it is clear that the Au adatoms preferentially occupy the narrow sites of the ($\sqrt2$~$\times$~$\sqrt2$)R45$^{\circ}$ reconstruction; occupation of the wide site was never observed. By analogy to the Fe-Fe$_3$O$_4$(001) surface \cite{ParkinsonFe}, for which the data are similar, we propose that the Au adatoms are two-fold coordinated to surface O atoms, i.e. in the site labeled “N” in Figure 1b. The high thermal stability and lack of RT diffusion suggests a deep well in the adsorption energy located at the narrow sites. 

It is surprising that geometrical relaxations of the order of 0.1 \AA{} in the clean surface could induce such a strong site preference. Furthermore, on purely geometric grounds one might expect the opposite preference, since an Au atom is significantly larger than an Fe atom, and the wide site would provide more space and allow the adatom to relax into the surface. For the clean surface DFT+U calculations \cite{Lodziana, Mula} predict charge/orbital order in the subsurface layers of the ($\sqrt2$~$\times$~$\sqrt2$)R45$^{\circ}$ reconstruction. In the first subsurface layer, charge ordered octahedral Fe(B) cations reside directly beneath the O atoms to which we propose the Au to bind. Specifically, pairs of Fe cations with Fe$^{3+}$-like and Fe$^{2+}$-like character are located beneath the wide and narrow sites, respectively, rendering the sites inequivalent from an electronic perspective. Thus, it is tempting to conclude that the templating property of the Fe$_3$O$_4$(001) surface is primarily driven by these subtle electronic effects.  

Charge transfer between metallic adatoms and the substrate is important for reactivity toward specific reactions. DFT calculations suggest that the partial vacation of the d-levels in positively charged Au adatoms reduces the CO adsorption energy and the activation barrier for CO oxidation \cite{Qiao, Fang}. Flynn and coworkers studied Au adatoms adsorbed at defects on Fe$_3$O$_4$(111) and concluded a positive charge state and demonstrated CO adsorption at low temperature \cite{Rim}. On ultrathin oxide films both positive and negative charging of adatoms has been demonstrated and attributed to electron transfer through the oxide \cite{Brown, Giordano, Nilius}. For positively charged adatoms a high work function metal support, an easily reducible oxide, and a dipole orientation favoring charge transfer into the support have been identified as important requirements \cite{Giordano}. Viewing Fe$_3$O$_4$(001) as a metallic bulk supporting a semiconducting polar ($\sqrt2$~$\times$~$\sqrt2$)R45$^{\circ}$ reconstruction, all of these requirements are met. Thus we consider it likely that the positive binding energy shift observed for Au in XPS (Figure 3) is representative of a positively charged Au adatom and not merely the result of final state effects \cite{Torelli}.

The Au-Fe$_3$O$_4$(001) system represents a unique model system for the study of the properties of oxide supported isolated adatoms. The key difference to prior model systems featuring Au adatoms is that the high density and thermal stability of adatoms will facilitate studies of their chemistry under realistic reaction conditions, allowing a resolution of their role in catalysis. Furthermore, the unambiguous nature of the site preference observed here will provide an excellent test case for the ability of theoretical calculations to correctly model this strongly correlated electron system. Finally, our observation that Fe adatoms show a similar site preference \cite{ParkinsonFe} suggests that the unusual templating properties of the Fe$_3$O$_4$(001)-($\sqrt2$~$\times$~$\sqrt2$)R45$^{\circ}$ surface may prove more universal, and that stable adatom arrays may be created for other catalytically active metals. 

Acknowledgement:  This material is based upon work supported as part of the Center for Atomic-Level Catalyst Design, an Energy Frontier Research Center funded by the U.S. Department of Energy, Office of Science, Office of Basic Energy Sciences under Award Number DE-SC0001058.  The authors acknowledge Prof. Z. Mao and T.J. Liu (Tulane University) for the synthetic sample used in this work and discussions with David Sholl and Thomas Manz (Georgia Tech.).

\end{document}